\documentclass[a4paper]{article}
\usepackage{graphics,amsmath,amsfonts}
\usepackage[sort&compress,numbers]{natbib}

\newcommand{\ket}[2]{|#1\rangle _{#2}}
\newcommand{\bra}[2]{\langle _{#1}#2|}

\title{Nonlocality and information flow: The approach of Deutsch and Hayden}

\author{C.G.\,Timpson\thanks{\tt christopher.timpson@queens.ox.ac.uk}\\ \textit{The Queen's College, Oxford University}}
\date{18 December 2003}

\begin{document}

\maketitle
\begin{abstract} 
\noindent Deutsch and Hayden claim to have provided an account of quantum mechanics which is particularly local, and which clarifies the nature of information transmission in entangled quantum systems. In this paper, a perspicuous description of their formalism is offered and their claim assessed. It proves essential to distinguish, as Deutsch and Hayden do not, between two ways of interpreting the formalism. On the first, conservative, interpretation, no benefits with respect to locality accrue that are not already available on either an Everettian or a statistical interpretation; and the conclusions regarding information flow are equivocal. The second, ontological, interpretation, offers a framework with the novel feature that global properties of quantum systems are reduced to local ones; but no conclusions follow concerning information flow in more standard quantum mechanics. 
\end{abstract}

\section{Introduction}

The existence of entanglement, and the associated questions concerning nonlocality, are of perennial interest in the foundations of quantum mechanics \cite{EPR,schrodinger1,bell1964,redheadbook,Maudlin:non-loc}. Latterly, following the development of quantum information theory \cite{nielsen:chuang}, entanglement assisted communication \cite{superdense,teleportation} has presented a new sphere in which puzzles may arise. In this context, an important development has been the claim of Deutsch and Hayden \cite{dh} to provide an especially local story about quantum mechanics by making use of the Heisenberg picture. Moreover, they claim finally to have clarified the nature of information flow in entangled quantum systems, reaching the conclusion that information is a local quantity, even in the presence of entanglement. The aim of this paper is to assess these claims.

Their discussion takes place within the context of unitary quantum mechanics without collapse, and without the addition of determinate values; and they proceed to make two claims to locality. First, they suggest, even in the presence of entanglement, the state of the global system can in fact be seen to be completely determined by the states of the individual subsystems, when these states are properly construed (a conclusion not available in the usual Schr\"odinger picture and one supposed to chime with Einstein's well-known demand for a \textit{real state} for spatially separated systems \cite{einsteinauto}). Second, the effects of local unitary operations, again, even in the presence of entanglement, are \textit{explicitly} seen to be local in their picture.   

However, before the implications of their formalism may be assessed, something needs to be said about how it is to be interpreted. Deutsch and Hayden are not explicit on this point and do not offer any interpretation. This proves problematic as two different modes of interpretation of their formalism may be discerned---what may be called the \textit{conservative} and the \textit{ontological} interpretations---and quite different conclusions follow concerning the questions of locality 
and information flow within these interpretations.

The conservative interpretation, perhaps the most natural way of reading the Deutsch-Hayden paper, takes the formalism at face value, simply as a re-writing of standard unitary quantum mechanics. In this case, we shall see, there are no novel gains with respect to locality and Deutsch and Hayden's claims about information flow prove at best misleading. Under the ontological interpretation, though, a dramatic departure from our usual ways of understanding quantum mechanics is made and a wholly new range of intrinsic properties of subsystems introduced. These would substantiate Deutsch and Hayden's claims, but at a certain cost of plausibility. We should note too that the ontological interpretation of the Deutsch-Hayden formalism is best seen as the postulation of a new type of theory, rather than being a new way of interpreting familiar quantum mechanics.   

The discussion will begin in Section~\ref{formalism}, where the machinery of the Deutsch-Hayden approach is outlined, in particular, the mathematics that lies behind the two claims to locality. These claims are then assessed (Section~{\ref{assess}}), for the conservative and ontological interpretations in turn. 

Note that in Deutsch-Hayden we have a formalism without collapse and without the addition of determinate values. If we are to consider the question of the locality of their approach, the appropriate comparisons are therefore with other approaches that are consistent with this assumption. One the one hand we should compare with a realist approach of the Everett stripe \cite{everett,simonrelativism,wallace}, while on the other we should compare with a form of statistical interpretation, by which I mean an interpretation in which quantum mechanics merely describes probabilities for measurement outcomes for ensembles, there is no description of individual systems and collapse does not correspond to any real physical process for individual systems. The question to be  answered, then, is: do Deutsch and Hayden present us with advantages with respect to locality that are not also shared by these other approaches? We shall see that under the conservative interpretation, they do not. 

In Section~\ref{info}, attention finally turns to the question of information flow in entangled systems. In Section~\ref{whereabouts} the nature of the question at issue is clarified, before  Deutsch and Hayden's explanation of quantum teleportation and their introduction of the concept of locally inaccessible information is considered (Section~\ref{explaining}). Their claims regarding the nature of information flow are then evaluated for the conservative and ontological interpretations in turn (Section~\ref{assessinfo}), along axes provided by three questions: i) Have Deutsch and Hayden finally given the correct account of teleportation, as compared to related accounts such as that of Braunstein \cite{braunstein:irreversible}? ii) Is the concept of locally inaccessible information useful? iii) Have they provided us with a new concept of information, or quantum information? We close with a brief summary.

\section{The Deutsch-Hayden Picture}\label{formalism}

Deutsch and Hayden consider a network of $n$ interacting two-state systems (\textit{qubits}, in the quantum computation literature) as their model of a general quantum system. They take as the object describing the state of the $i$th qubit at time $t$ a triple \[\mathbf{q}_{i}(t)=(q_{i,x}(t),q_{i,y}(t),q_{i,z}(t))\] of $2^{n}\times2^{n}$ Heisenberg picture operators satisfying the familiar commutation and anti-commutation relations of the Pauli spin operators. To see how this representation works, let us first recall the basics of the Heisenberg picture. 

\begin{sloppypar}
As expressed in the equations
\begin{equation}
\bra{}{\psi(t)}A\ket{\psi(t)}{}=\bra{}{\psi}U^{\dagger}AU\ket{\psi}{}=\bra{}{\psi}A(t)\ket{\psi}{},
\end{equation}
time dependence in quantum mechanics can either be associated with the vector (\textit{ket}) representing the state, or with the operator representing the observable. In the Schr\"odinger picture, the state ket undergoes unitary evolution ($\ket{\psi}{} \mapsto U\ket{\psi}{}$); in the Heisenberg picture, the state ket remains unchanged and the basis kets $\{\ket{\alpha_{i}}{}\}$ of the Hilbert space are evolved ($\ket{\alpha_{i}}{} \mapsto U^{\dagger}\ket{\alpha_{i}}{}$).
Another useful way of representing these facts is as follows.
\end{sloppypar}

It is well known (e.g. \cite{fano}) that the set of $N\times N$ complex Hermitian matrices forms an $N^{2}$ dimensional real vector space, $V_{h}(C^{N})$, if we define an inner product $(A,B)=\mathrm{Tr}(AB),\; A,B\in V_{h}(C^{N})$ and norm $||A||=\sqrt{\mathrm{Tr}A^2}$. Just as in our familiar examples of vector spaces, e.g. Euclidean ${\mathbb R}^3$, it is useful to define a set of basis vectors for the space. We require $N^{2}$ linearly independent operators $\Gamma_{j} \in V_{h}(C^N)$, and it may be useful to require orthogonality: $\mathrm{Tr}(\Gamma_{j}\Gamma_{j^{\prime}})=\mathrm{const.}\delta_{jj^{\prime}}$.

Any observable can then be represented in this space in the form:
\begin{equation}
\mathbf{A}=\sum_{j=0}^{N^2-1}\mathrm{Tr}(A\Gamma_{j})\Gamma_{j}=\sum_{j=0}^{N^2-1}A_{j}\Gamma_{j},
\end{equation}
where the $\mathrm{Tr}(A\Gamma_{j})=A_{j}$ are the components of the vector $\mathbf{A}$ representing the observable $A$. The density matrix $\rho$ can also be written as a vector:
\begin{equation}\label{density matrix}
{\varrho}=\frac{\mathbf{1}}{N} + \sum_{j=1}^{N^{2}-1}\mathrm{Tr}(\rho \Gamma_{j})\Gamma_{j}=\sum_{j=0}^{N^2-1}\rho_{j}\Gamma_{j},
\end{equation}
where $\Gamma_{0}$ has been chosen as $\mathbf{1}$, the identity.
In this representation, the expectation value of $A$ is just the projection of the vector ${\varrho}$ onto the vector $\mathbf{A}$: $\langle A \rangle_{\rho}= \mathrm{Tr}(A\rho)=(\mathbf{A}.{\varrho}$).
The equivalence between the Schr\"odinger and Heisenberg pictures now takes on a very graphic form. We can either picture leaving the basis vectors (operators) as they are and rotating the vector ${\varrho}$ under time evolution, or we can picture rotating the basis vectors (and hence any observable $A$) in the opposite sense, and leaving ${\varrho}$ unchanged. In either case, the angle between the two resulting vectors and hence the expectation value is clearly the same: $\mathbf{A}(t).{\varrho}=\mathbf{A}.{\varrho}(t)$.

Writing the time dependence out explicitly, we will have, in the Heisenberg picture:
\begin{equation}
\mathbf{A}(t)=\sum_{j}A_{j}U^{\dag}(t)\Gamma_{j}U(t), \label{A(t)}
\end{equation}
while in the Schr\"odinger picture,
\begin{equation}
\varrho(t)=\sum_{j}\mathrm{Tr}(\rho U^{\dag}(t)\Gamma_{j}U(t))\Gamma_{j} = \sum_{j}\langle\Gamma_{j}(t)\rangle_{\rho}\Gamma_{j}. \label{rho(t)}
\end{equation}
The expectation value of observable $A$ at time $t$ is simply $\sum_{j}A_{j}\langle\Gamma_{j}(t)\rangle_{\rho}$. 

Notice that in both expressions (\ref{A(t)}) and (\ref{rho(t)}), the time evolved operators $\Gamma_{j}(t)= U^{\dag}(t)\Gamma_{j}U(t)$ feature. These operators, along with their expectation values $\langle\Gamma_{j}(t)\rangle_{\rho}$, will be our main objects of interest.

What should we choose as basis vectors? For $N=2$, the set of Pauli operators forms an orthogonal basis set, $\mathrm{Tr}(\sigma_{i}\sigma_{j})=2\delta_{ij}$, (we adopt the convention that $\sigma_{0}$ denotes the identity) thus we can choose $\sqrt{2}\,\Gamma_{j}\in\{\mathbf{1},\sigma_{x},\sigma_{y},\sigma_{z}\}$ to provide an orthonormal basis $\{\Gamma_{j}\}$.\footnote{The choice of the Pauli operators as a basis set gives us the familiar Bloch sphere representation of the density matrix of a two-state system.} We are then interested in the behaviour of the set $\{U^{\dag}(t)\,(\sigma_{i}/\sqrt{2})\,U(t)\}$.

So far, all we have done is translate some very familiar results into the language of the space $V_{h}(C^{N})$. We now make the all-important move that provides the core result of the Deutsch-Hayden picture (following Gottesman~\cite{gottesman}). That is, we note that unitary transformations of operators have the property of being a multiplicative group homomorphism\footnote{A map $f:\mathcal{A}\mapsto \mathcal{B}$ is a \textit{group homomorphism} if $\forall a_{1},a_{2}\in \mathcal{A}, f(a_{1}a_{2})=f(a_{1})f(a_{2})$.}:
\begin{equation} U^{\dag}ABU=(U^{\dag}AU)(U^{\dag}BU). \end{equation}
In other words, the time evolution of a product will be given by the product of the time evolution of the individual operators. Thus we do not need to follow the evolution of the whole basis set of operators, but only of a generating set. For example, in the $N=2$ case, noting that $\sigma_{x}\sigma_{y}=i\sigma_{z}$, we see that $\sigma_{z}(t)=-i\sigma_{x}(t)\sigma_{y}(t)$ and that we need only follow the evolution of the generating set $\{\sigma_{x},\sigma_{y}\}$ to capture the time evolution of the whole system. (For completeness, note that $\sigma_{i}^{2}=\mathbf{1}$; the time evolution of the identity is of course trivial.)  

For $N=2^{n}$, $n$-fold tensor products of Pauli matrices will provide us with an orthogonal set, thus our basis operators will be
\begin{equation}\label{gammaj}
\Gamma_{j}=\frac{1}{\sqrt{2^{n}}}\:\sigma^{1}_{m_{1}}\otimes\sigma^{2}_{m_{2}}\otimes\ldots\otimes\sigma^{n}_{m_{n}};
\end{equation}
where the index $j$ runs from 0 to $(4^{n}-1)$ and labels an ordered $n$-tuple $<\!\!m_{1},m_{2},\ldots,m_{n}\!\!>$, $m_{i}\in\{0,1,2,3\}$. We are interested in the behaviour of the $4^{n}$ $\Gamma_{j}(t)$; again, however, we need only track the evolution of objects of the form
\[ \mathbf{1}\otimes\mathbf{1}\otimes\ldots\otimes\sigma^{i}_{m_{i}}\otimes\ldots\otimes\mathbf{1}, \]
which we denote $q_{i,m_{i}}$; the $\Gamma_{j}$ are given by ordinary matrix multiplication of these objects:
\begin{equation}
\Gamma_{j}=\prod_{i=1}^{n}\frac{1}{\sqrt{2}}\:q_{i,m_{i}}\,.
\end{equation}
The behaviour of the $\Gamma_{j}(t)$ is thus determined by following the time evolution of a minimum of $2n$ of the $q_{i,m_{i}}$ and taking appropriate products. 

The $q_{i,m_{i}}$ with $m_{i}$ running from 1 to 3 are, of course, the components of the Deutsch-Hayden descriptor $\mathbf{q}_{i}$. This choice of three operators per system as the basic objects whose time evolution we are to follow is more than is strictly necessary for a generating set, but it leads to a very simple description of the properties of an individual system, as we shall shortly see. First, however note that
the density matrix at time $t$ can now be written as
\begin{equation}
{\varrho}(t)=\frac{1}{2^n}\sum_{m_{1}m_{2}\ldots m_{n}}\biggl\langle \prod_{i}q_{i,m_{i}}(t)\biggr\rangle_{\!\!\!\rho}\prod_{i}q_{i,m_{i}}\,. \label{density}
\end{equation} 
That is, the $4^{n}$ components $\rho_j(t)$ of the vector representing the density matrix at time $t$ are given by the expectation values of products of the $q_{i,m_{i}}(t)$. The state of the joint system at time $t$ is thus completely determined by the time evolution of the $2n$ or $3n$ chosen $q_{i,m_{i}}$ and the initial state $\rho$. To see the significance of the triple $\mathbf{q}_{i}$, note that any observable on the $i$th system alone will have the form:
\begin{equation}
A^{i}=\sum_{m_{i}=0}^{3}A_{m_{i}}(\mathbf{1}\otimes\mathbf{1}\otimes\ldots\otimes\sigma^{i}_{m_{i}}\otimes\ldots\otimes\mathbf{1})=A_{0}\mathbf{1}^{\otimes n}+\sum_{m_{i}=1}^{3}A_{m_{i}}q_{i,m_{i}},
\end{equation}
thus $\mathbf{q}_{i}(t)$ tells about observables on the $i$th system at time $t$ and $\langle \mathbf{q}_{i}(t) \rangle_{\rho}$ determines their expectation values. Equivalently, the three components of $\langle \mathbf{q}_{i}(t)\rangle_{\rho}$ give us the interesting components of the vector ${\varrho}(t)$ lying in the subspace spanned by observables pertaining to the $i$th system alone; and with renormalisation, the components, in our vector representation, of the reduced density matrix of the $i$th system.

Explicitly, this reduced density matrix is:
\begin{equation}
\rho^{i}(t)=\frac{1}{2}\sum_{m_{i}}\langle q_{i,m_{i}}(t)\rangle_{\rho}\,\sigma^{i}_{m_{i}}.\label{rdmi}
\end{equation}
It is also easy to write down the reduced density matrix for any \textit{grouping} of subsystems. If we were interested in the systems $i$, $j$ and $k$, say, taking the partial trace of (\ref{density}) over the other systems will give us a reduced state of the form:
\begin{equation} 
\rho^{ijk}(t)=\frac{1}{8}\sum_{m_{i}m_{j}m_{k}}\langle q_{i,m_{i}}(t)q_{j,m_{j}}(t)q_{k,m_{k}}(t)\rangle_{\rho}\,\sigma^{i}_{m_{i}}\otimes\sigma^{j}_{m_{j}}\otimes\sigma^{k}_{m_{k}}.\label{rdmijk}
\end{equation}

So we have now seen the basis for the first claim to locality: given just the descriptors $\mathbf{q}_{i}(t)$ for each individual system, and the initial state $\rho$, we may calculate the reduced density matrix for each subsystem, \textit{and} the density matrix for successively larger groups of subsystems, up to and including the density matrix for the system as a whole.

We may note in passing another interesting feature of the Deutsch-Hayden formalism. A question that often arises, particularly in discussion of quantum correlations, is whether different preparations of the same density matrix really correspond to physically distinct situations, as all observable properties of systems having the same density matrix are identical. A pleasing aspect of the Deutsch-Hayden set-up is that it provides a representation in which differences in the way systems are prepared may find direct expression in the formalism\footnote{Although, it must be noted that as we are in the context of no-collapse quantum mechanics, the possibility does not obtain of preparing a distant system in a particular way via collapse, \textit{\`a la} EPR.}. For example, it may be the case that $\langle \mathbf{q}_{i}(t)\rangle_{\rho}=\langle \mathbf{q}_{j}(t)\rangle_{\rho}$ i.e. the two systems have the same reduced density matrix, but that $\mathbf{q}_{i}(t)$ and $\mathbf{q}_{j}(t)$ differ, representing differences in their histories.

\subsection{Locality claim (2): Contiguity}

Let us now consider the second claim to locality. This, recall, was the claim that it can be seen explicitly in the Deutsch-Hayden formalism that local unitary operations have only a local effect. 
As Jozsa \cite{jozsa} has emphasized, this aspect of the Deutsch-Hayden picture is in fact a re-expression of the no-signalling theorem.

\begin{figure}
\scalebox{0.75}{\includegraphics{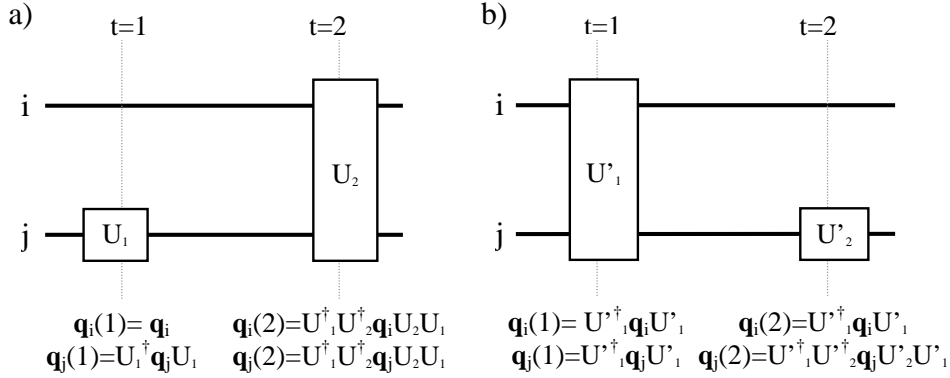}}
\caption{a) At $t=1$, a unitary operation, $U_{1}$, which acts only on system $j$ is applied; the descriptor of system $i$, $\mathbf{q}_{i}(1)$, is unaffected. After $i$ and $j$ interact via $U_{2}$ at $t=2$, however, $\mathbf{q}_{i}(2)$ will depend on the operation $U_{1}$. In (b) systems $i$ and $j$ initially interact via $U^{\prime}_{1}$. At $t=2$, $U^{\prime}_{2}$, acting on $j$ alone, is applied; $\mathbf{q}_{i}(2)$ is unaffected.
}\label{contiguity}
\end{figure}

In the Heisenberg picture, a sketch of a simple version of the theorem would be as follows: let us write an observable acting on subsystem $i$ alone as $A^{i}=\mathbf{1}\otimes A$; at time $t$, $A^{i}(t)=U^{\dag}(t)(\mathbf{1}\otimes A)U(t)$. Suppose $U(t)$ does not act on $i$, then $A^{i}(t)=(U^{\dag}\otimes\mathbf{1})(\mathbf{1}\otimes A)(U\otimes\mathbf{1})=\mathbf{1}\otimes A$, i.e. an observable is unaffected by unitary operations on systems it does not pertain to. Now consider our $q_{i,m_{i}}$; the foregoing clearly applies to them --- a unitary operation on a system $j$ does not affect $q_{i,m_{i}}$. More generally, if our network of $n$ systems were divided up into two subsets of systems, $M$ and $N$, whose members interact amongst themselves but not with systems from the other subset, then the unitary operator describing the time evolution of the network will factorise: $U^{M}\otimes U^{N}$. Then the $q_{i,m_{i}}$ for $i\in M$ will not be affected by $U^{N}$, nor those for $i \in N$ by $U^{M}$. We can do more than merely note that the descriptors of a set of interacting systems do not depend on unitary operations on a disjoint set, however. In fact we can see that the descriptor at time $t$ of a given system will depend, apart from the history of operations applied to it alone, only on its previous interactions and on the histories and past interactions of the systems it has interacted with. This property may be called \textit{contiguity}; and is best seen with a simple example (Fig~\ref{contiguity}). 

Imagine we have two systems, $i$ and $j$ and that we are going to perform two unitary operations. First, at $t=1$, we perform $U_{1}$, which acts on $j$ alone; clearly, after this operation, $q_{i,m_{i}}(1)=U^{\dag}_{1}q_{i,m_{i}}U_{1}=q_{i,m_{i}}$. Next we allow $i$ and $j$ to interact via $U_2$; now, however, $q_{i,m_{i}}(2)=U^{\dag}_{1}U^{\dag}_{2}q_{i,m_{i}}U_{2}U_{1}$. Because $U_{2}$ acts on both $i$ and $j$, $U_{1}$ no longer factors out; interaction causes the $q_{i,m_{i}}$ to lose the form of a product of a single Pauli operator with the identity and they can pick up a dependence on what has happened to the system that $i$ has interacted with. We can say that all this remains happily local, however, as this dependence on the history of $j$ only arises following an entangling interaction between the two systems. The reasoning extends in the obvious way to more complicated chains; if $j$ had previously interacted with $k$, then once $i$ and $j$ interact, the $q_{i,m_{i}}(t)$ pick up what they would not previously have had, a dependence on what has happened to $k$; and so on.

To re-emphasize that the Deutsch-Hayden descriptor of a system at time $t$ will not, however, depend on what happens at $t$ to a system with which it has interacted in the past, we take the following simple example (Fig~\ref{contiguity}). Again consider two systems $i$ and $j$; this time, however, we begin by allowing them to interact via a unitary operation $U^{\prime}_{1}$, then 
\begin{eqnarray}
q_{i,m_{i}}(1) & = & U^{\prime\dag}_{1}q_{i,m_{i}}U^{\prime}_{1}\neq q_{i,m_{i}},\;\; \mathrm{and} \nonumber\\
q_{j,m_{j}}(1) & = & U^{\prime\dag}_{1}q_{j,m_{j}}U^{\prime}_{1}\neq q_{j,m_{j}}. 
\end{eqnarray}
\sloppy Now we perform $U^{\prime}_{2}$, which acts on $j$ alone.
Whilst $q_{j,m_{j}}(2)= U^{\prime\dag}_{1}U^{\prime\dag}_{2}q_{j,m_{j}}U^{\prime}_{2} U^{\prime}_{1}$, for the descriptor of $i$ we have
\begin{equation} q_{i,m_{i}}(2)= U^{\prime\dag}_{1}U^{\prime\dag}_{2}q_{i,m_{i}}U^{\prime}_{2} U^{\prime}_{1}=U^{\prime\dag}_{1}q_{i,m_{i}}U^{\prime}_{1},\end{equation} 
$U^{\prime}_{2}$ factors out; there is no immediate dependence on what happens at the present only to $j$, even when $i$ and $j$ have interacted in the past.

\begin{figure}\begin{center}
\scalebox{0.5}{\includegraphics{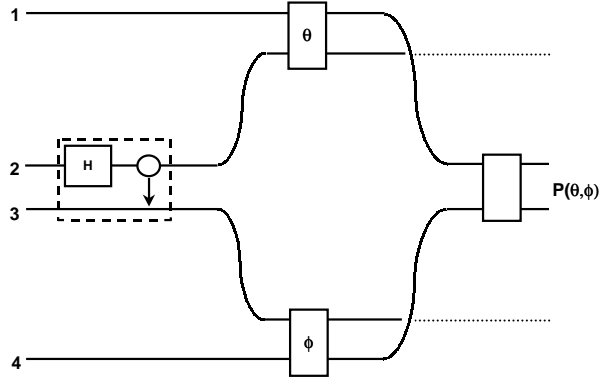}}\end{center}
\caption{A Bell experiment. An entangled state of systems 2 and 3 is prepared (here by the action of a Hadamard gate, H, followed by a controlled-NOT operation---the circle indicates the control qubit, the point of the arrow, the target) and shared between two distant locations. A measurement at an angle $\theta$ is performed on 2 and the outcome recorded in system 1; a measurement at an angle $\phi$ made on 3 and recorded in 4. Time runs along the horizontal axis. Note that in no collapse quantum mechanics without added values, correlations do not obtain until they are displayed by a suitable joint measurement. }\label{Bell}
\end{figure}

The picture, then, is that following an interaction, the descriptor of a system $i$ picks up a backwards looking (and hence what we might call a local, or contiguous) dependence on what has happened to the system that $i$ has interacted with, and on the previous interactions of that system. 
As an illustration, let us consider how the non-factorisable probability distributions for Bell-type experiments come about in this formalism (Fig.~\ref{Bell}).

As usual, we begin by preparing a pair of systems (2 and 3) in an entangled state. These systems are spatially separated and two local measurements performed, at an angle $\theta$ on system 2 and an angle $\phi$ on system 3. The outcomes are recorded into systems 1 and 4 respectively. Immediately following the measurement, the descriptor of system 1 will depend on $\theta$, but not on the parameter characterizing the distant measurement, $\phi$. However, as system 1 has interacted with system 2, its descriptor will \textit{also} depend on what has happened to 2 in the past; which was, in this case, an entangling interaction between 2 and 3. Similarly, the descriptor of 4 following the local measurement will depend on $\phi$ and not on $\theta$, but will depend too on what happened to system 3 --- that is, on 3's initial entangling with 2. Because the descriptors of 1 and 4 depend, following the pair of local measurements, on the initial entangling interaction between 2 and 3, their product can give rise to the familiar non-factorisable probability distribution when 1 and 4 are subsequently brought together and joint measurements performed.      

It is tempting to think of the contiguity property of the Deutsch-Hayden descriptors 
as depicting a causal chain in which dependence on the parameters characterising the history of a system is passed on during interactions, or even more metaphorically, in terms of information about the relevant history of a system being transmitted via local interactions. More soberly, we see that if the $\mathbf{q}_{i}$ are taken to be the primary objects of interest then the effects of local unitary operations on these are indeed explicitly seen to be local, as the descriptor of a system cannot come to depend on a parameter characterising a unitary operation selected in a distant region without the system having undergone an appropriate chain of \textit{local} interactions.  
As we have said, however, this is just the no-signalling theorem writ large.

\section{Assessing the Claims to Locality}\label{assess}

Having outlined the machinery of the Deutsch-Hayden approach, we may now consider the status of its claim to provide a particularly local picture of quantum mechanics. As remarked in the introduction, it is necessary to distinguish two modes of interpretation of the formalism.

\subsection{The Conservative Interpretation}

The \textit{conservative interpretation} is to take the formalism at face-value, simply as a re-writing of standard (unitary) quantum mechanics, in which we fix the initial state $\rho$ and track time evolution via the $\mathbf{q}_{i}(t)$. If we want to talk in terms of properties, we may see the $\mathbf{q}_{i}(t)$, against the background of a chosen $\rho$, as denoting propensities for the display of certain individual and joint probability distributions for measurement outcomes, via equations (\ref{rdmi}) and (\ref{rdmijk}).    

\subsubsection{Locality Claim (1)}\label{loc1}

The first claim to locality was that the global state can be seen to be determined by the states of individual subsystems. What is certainly true is that given the $n$ $\mathbf{q}_{i}(t)$, the $4^{n}$ $\Gamma_{j}(t)$ are determined and hence we can keep track of the changes to the joint system over time. Note, however, that the initial global state $\rho$ still has to be specified and plays a very important role. It is needed to determine the experimentally accessible properties of individual and joint systems; both the $\Gamma_{j}(t)$ \textit{and} $\rho$ are required to determine expectation values of measurements. That it is the \textit{global} state is crucial, as in general in the presence of entanglement, $\langle q_{i,m_{i}}\!(t)\,q_{j,m_{j}}\!(t)\rangle_{\rho}\neq \langle q_{i,m_{i}}(t)\rangle_{\rho}\langle q_{j,m_{j}}(t)\rangle_{\rho}$. 

With the global state of the system still playing such an important role, however, it is not clear that we have yet gained much in the way of locality by considering the Deutsch-Hayden construction under the conservative interpretation.
Taking the simplest picture of a time evolving density operator, products of the $\mathbf{q}_{i}(t)$ determine how \textit{any} given initial state will evolve; it is no surprise if the initial state of the joint system is specified and we have kept track of the changes to the system (albeit that these are fixed by the individual $\mathbf{q}_{i}(t)$) that we then know what the final state will be.

In reply it is open to Deutsch and Hayden to argue that appeal to the global state is in fact innocuous, as a standard initial state can always be chosen and the $\mathbf{q}_{i}(0)$ adjusted accordingly. To be sustained, however, this line of argument commits one to the ontological interpretation, which we shall consider in due course. For now, let us consider the status of the second locality claim under the conservative interpretation.

\subsubsection{Locality Claim (2)}

We begin by asking why it might seem important to show explicitly that local unitary operations have only a local effect. (We recall, of course, that the standard no-signalling theorem already assures us that local unitaries will not have any effect on the probability distributions for distant measurements). It is clear that if we were only to consider the question of nonlocality as it is usually raised in the context of Bell-type experiments, then the Deutsch-Hayden approach would not offer us any distinctive advantages. For, as mentioned in the Introduction, their point of departure is to assume no-collapse quantum mechanics with no determinate values added, thus the appropriate comparisons must either be with an Everettian or a statistical interpretation. But it is well known that the Everett interpretation does not suffer from the familiar difficulties with nonlocality in the Bell or EPR setting that accrue to theories involving collapse or additional variables (indeed, this is often presented as one of the selling-points of the approach); while for a statistical interpretation, the familiar no-signalling theorem does all that could be required to ensure that nonlocality does not arise (see \cite{erpart1} for further discussion and references). Thus if one is considering the question of locality in this context, the crucial factor is the assumption of quantum mechanics without a real process of collapse, and without additional variables, rather than anything distinctive about the Deutsch-Hayden approach.

However, things may look rather less clear-cut when one considers the phenomena of entanglement assisted communication such as superdense coding \cite{superdense} and teleportation \cite{teleportation}. These phenomena vividly illustrate the fact that in the presence of entanglement, local unitary operations can have a very significant effect on the \textit{global} state of the system.
And might this not indicate a novel sort of nonlocality of which even the Everett interpretation would be guilty? 
If so, the Deutsch-Hayden approach would seem to offer a clear advantage, with its explicit locality regarding the effects of local unitary operations.

\begin{figure}
\begin{center}\scalebox{0.7}{\includegraphics{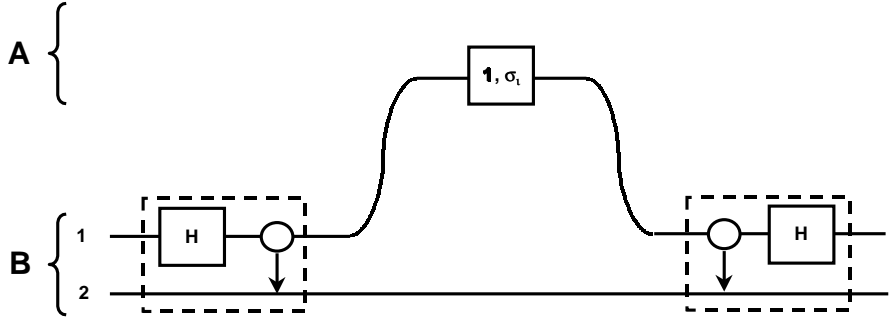}}\end{center}
\caption{Superdense coding. A maximally entangled state of systems 1 and 2 is prepared by Bob (B). System 1 is sent to Alice (A) who may do nothing, or perform one of the Pauli operations. On return of system 1, Bob performs a measurement in the Bell basis, here by applying a controlled-NOT operation, followed by the Hadamard gate. This allows him to infer which operation was performed by Alice.}\label{sdcoding}
\end{figure}

Consider the example of superdense coding in more detail (Fig.~\ref{sdcoding}).  
In this protocol, Alice is able to send Bob two bits of information with the transmission of a \textit{single} qubit, by making use of the global effect of a local operation. 

The two parties begin by sharing a maximally entangled state; let us say the singlet state. Then, simply by applying one of the Pauli operators to her half of the shared system, Alice may flip the joint state into one of the others of the four orthogonal, maximally entangled Bell states: a local operation has resulted in a change in the global state that is as great as could be --- from the initial state to one orthogonal to it. Now, if Alice sends her half of the shared system to Bob, he just needs to measure in the Bell basis to determine which of the four operations Alice performed, arriving at two bits of information. In this protocol, the possibility of changing the global state by a local operation has been used to send information in a very unexpected way. The phenomenon of teleportation may also be viewed as arising from the fact that the set of maximally entangled states may be spanned by local unitary operations \cite{braunstein:twist}.

So, does the example of entanglement assisted communication indicate an important sphere in which Deutsch-Hayden presents benefits of locality? Note that these examples do not affect the question of locality for the statistical interpretation, as on this interpretation the quantum state does not correspond to anything real. But one might be interested in a more realist approach. Thus we should ask how the Everett interpretation fares with locality in entanglement assisted communiction.   

It can in fact be argued that the examples of superdense coding and teleportation do not demonstrate a new form of nonlocality in Everett \cite{erpart1}. Our worry is about the effect on the global state of local operations; however, even if we are being robustly realist about it, the global state is not itself a spatio-temporal entity. Thus changes in the global state do not correspond to local \textit{or} to non-local changes. It is better to think in terms of changes to properties of the systems; but it is clear that unlike the sort of change that would be associated with collapse, the effects of local unitary operations that we are considering do not give rise to any changes in local and non-relational properties of the separated systems (i.e., locally observable probability distributions are unchanged). 
Thus, although certainly striking, and non-classical, the potential global effects of a local unitary operation in the presence of entanglement are not appropriately construed as non-local. 

The case is clear enough for superdense coding; teleportation invites a further brief comment. In the standard teleportation scenario, Alice and Bob again share a maximally entangled state. Alice's aim is now to send Bob an unknown quantum state, rather than some classical information. She does this by performing a Bell-basis measurement on her half of the entangled state and the system whose state she is trying to transmit. This has the effect of flipping Bob's half of the entangled pair into a state that differs from the unknown state by one of four unitary operations, depending on the outcome of Alice's measurement. If the outcome of Alice's measurement is then transmitted to Bob, he may perform the appropriate conditional operation to attain a system in the state that Alice was trying to send. When this protocol is analysed from the Everett perspective, the significant feature is that immediately Alice performs her measurement, and before she sends a record of the outcome to Bob, Bob's system will already have acquired a definite state related to the state Alice is sending, relative to the outcome of her measurement. Which may make us worry again that there must be some nonlocality involved.

This feeling is quickly dispelled, though, when we note that what have changed as a result of Alice's measurement are the \textit{relative} states of Bob's system. This means that we are talking about relational properties of his system. But the fact that a relational property may change following an operation on one of the \textit{relata} does not imply a non-local effect\footnote{Consider the following classical example: We have two heaps of sand, $x$ and $y$, piled on the ground, some distance apart. Let us say $x$ is heavier than $y$. By adding a few more shovel-fulls to $y$, we may make this statement false; but this does not imply a non-local effect on $x$.}, as indeed these changes in the relative states do not. This claim is further elaborted in \cite{erpart1}; Vaidman \cite{vaidman} has also argued to the effect that teleportation does not involve nonlocality, when understood in Everettian terms.

The conclusion is that when considered under the conservative interpretation, the explicit locality in the effect of local unitary operations that the Deutsch-Hayden formalism provides in the contiguity of changes in the $\mathbf{q}_{i}(t)$ does not vouchsafe an important sense of locality that would be lacking in an Everettian or statistical interpretation. 
Indeed we can see that it would necessarily be quite misleading to suggest that the contiguity property points to a novel feature of locality in the Deutsch-Hayden formalism interpreted conservatively. As we have noted, the novelty must be supposed to concern the absence of any effect on the global state from local unitary operations, even in the presence of entanglement; and this indeed follows, in a trivial sense, if we \textit{fix} the initial state $\rho$ and track time evolution via the $\mathbf{q}_{i}(t)$, adopting the Heisenberg viewpoint. But what we described in the Schr\"odinger picture as a change in the global state following a local operation now merely becomes, in the Heisenberg picture, a change in the expectation values for some joint observables that can't be understood in terms of changes in expectation values for observables pertaining to subsystems. But why, if we were supposed to be worried at all, should we be less worried by changes in these joint expectation values as a result of local unitary operations, than in changes to the global state?

\subsection{The Ontological Interpretation}

Maudlin, in the course of his careful discussion of the question of holism in quantum mechanics, arrives at the following dialectical position:
\begin{quote}
We now have a reasonably clear question: according to the quantum theory, can the physical state of a system be completely specified by the attribution of physical states to the spatial parts of the system, together with facts about how those parts are spatiotemporally related? \cite[p.50]{maudlin}
\end{quote}
In standard quantum theory, the answer, of course, is \textit{no}. The point of the Deutsch-Hayden approach under the ontological interpretation is to answer instead `yes'.

To see how this might be achieved, recall why the conservative interpretation must fail to give an affirmative answer to Maudlin's question.

In the conservative interpretation, the assignment of properties at a given time is necessarily a joint venture between the global state $\rho$ and the descriptors; and as we noted (Section~\ref{loc1}), appeal \textit{has} to be made to global properties of the state. The $\mathbf{q}_{i}(t)$ cannot themselves be said to denote properties of the subsystems, rather, they determine what the effects of dynamical evolution would be for any possible initial state of the whole system. It is only when some particular initial state is specified that we may begin to talk about the properties of subsystems and of the whole; denoted by expectation values of the $\mathbf{q}_{i}(t)$ and products of the $q_{i,m_{i}}(t)$, respectively. And we have already noted a crucial feature several times: in general, the properties that are assigned to joint systems (expectation values for joint observables, or propensities for the display of certain joint probability distributions on measurement), will not be reducible to properties assigned to subsystems (individual expectation values and propensities). 

The ontological interpretation departs from this in two ways. First, the status of the global quantum state is fundamentally revised. A fixed standard state is adopted by convention (for example, the computational basis state $\ket{0}{}\ket{0}{}\ldots\ket{0}{}$) and it is delegated to playing a purely mathematical r\^ole in the machinery of the theory, rather than representing any physical contingency. Its status is now simply that of a \textit{rule} for reading off the observable properties of systems. Secondly, the $\mathbf{q}_{i}(t)$ are taken to represent intrinsic (i.e., non-relational) and occurrent (i.e., non-dispositional) properties of individual subsystems. The first feature is required of these properties if the global properties of the total system are to be reduced to the properties currently possessed by its subsystems; the second feature is a natural requirement in this context. A change in the descriptor of a system now represents a change in the actually possessed, intrinsic properties of the system. These intrinsic properties are clearly of a new sort; and they do not receive any further characterisation or explanation than is provided by their r\^ole in the formalism. Thus on the ontological interpretation, the content of the first claim to locality is that the global properties of the joint system are reducible to local, intrinsic properties of subsystems, while the content of the second is that \textit{changes} in the global properties are reducible to changes in the currently possessed properties of subsystems. Under the ontological intepretation, then, we certainly have an interesting thesis. Note that now, as adumbrated earlier, changes in the initial conditions of a system may be reflected in changes in the $\mathbf{q}_{i}(0)$, whereas under the conservative interpretation they would be represented by changes in the time-zero density matrix, $\rho(0)$\footnote{A half-way house is unsatisfactory. One might adopt a conventional fixed initial state in the conservative interpretation and adjust the $\mathbf{q}_{i}(0)$ accordingly, but this would not eliminate the global r\^ole of the state in determining joint properties, i.e. we do not have reducibility to individual properties, as in this interpretation the $\mathbf{q}_{i}(t)$ do not represent intrinsic properties.}.

It can hardly be emphasized enough that the approach of the ontological interpretation marks a considerable departure from our usual ways of thinking about quantum mechanics. Indeed it might best be thought of as the proposal of a new theory, in which the behaviour of the intrinsic properties denoted by the $\mathbf{q}_{i}(t)$ is fundamental\footnote{Note, however, that the ontological interpretation of Deutsch-Hayden lacks a measurement theory. Although we have a prescription for what the probability distributions associated with various measurements will be, we do not yet have a description of the measurement process itself, or of the obtaining of various outcomes, in terms internal to the theory. It might be thought that some sort of Everettian approach could be adopted, but as the relative state finds no place in the Deutsch-Hayden framework, it appears, at least \textit{prima facie}, to be resistant to standard Everettian analysis.}.  

In gaining with respect to reducibility, however, the ontological interpretation acquires what might be felt to be some rather objectionable features. The first is a problem of underdetermination.

The central, distinctive, claim of the ontological interpretation is that the intrinsic properties of a subsystem, denoted by the descriptor $\mathbf{q}_{i}(t)$ are fundamental. This means that there is a fact about which properties a given system actually possesses at any stage; and thus also, a fact about what the true descriptor of the system is. However the interpretation also involves a strict distinction between observable and unobservable properties. The observable properties are those that are given by expectation values. But this means that we can never in fact know the true descriptor of a system. We only have empirical access to expectation values and to the density matrices of systems, but continuously many different $\mathbf{q}_{i}(t)$ will be compatible with this data. The true descriptor of a system could be any one of the many that would provide consistency with both the density matrix of the subsystem (eqn.~(\ref{rdmi})) and that of the total system (eqn.~(\ref{density})). Thus the facts about the true descriptors; and hence about the intrinsic properties that systems actually possess, although supposedly the fundamental reality, are empirically inaccessible. According to the ontological interpretation, there is an important fact about what the correct descriptors of a set of systems are, but any assignment of descriptors to such a set will necessarily be underdetermined by the accessible data.

As a corollary of this point, it is worth remarking that the analogy Deutsch and Hayden suggest between their descriptors and Einstein's desired `real state' for separated systems might be overstated. While it may be the case that under the ontological interpretation, subsystems do indeed possess independent real states, we would still face the epistemological problem that this real state could never be determined by local measurements --- we could at most only ever learn the $\langle \mathbf{q}_{i}(t)\rangle_{\rho}$ for a system, when presented with a sufficient number of identically prepared systems.    
 
The second difficulty for the ontological interpretation, and one closely related to the underdetermination problem, is that the shift in meaning of the $\mathbf{q}_{i}(t)$, from determining time evolution for any given initial state, to denoting intrinsic properties of subsystems, induces a worrisome redundancy. In the normal quantum mechanical picture one can think of the $\mathbf{q}_{i}(t)$ in the following way.

Take some fixed sequence of unitary operations performed on a group of systems. This sequence will correspond to some particular evolution of the set of $\mathbf{q}_{i}(t)$. Now we could consider different initial quantum states for the set of systems; these states would evolve variously under the sequence of unitary operations whose effect is captured in the evolving $\mathbf{q}_{i}(t)$. 
At any given time, the actual quantum state of our group of systems could be one from a whole range, depending on which initial state was in fact chosen. The evolution of some particular initial state from time 0 to time $t$ may therefore be said to depict one history from the range of possible ones. To use the term favoured by philosophers, the evolution of this state represents the history of one possible world. A choice of different initial state is a choice of different possible world. 

Now the $\mathbf{q}_{i}(t)$ capture the effects of our sequence of unitary operations for all initial states. Thus their time evolution can be said to depict the histories of the \textit{entire set} of possible worlds; whilst the world from amongst these that is realised is determined by which initial state is chosen. However, when we move to the ontological view, the \textit{very same structure} (the sequence of time evolving $\mathbf{q}_{i}(t)$) only represents a \textit{single} world, as the choice of initial state is a fixed part of the formalism. What seems like it can represent a range of possible worlds, we are to suppose, can only represent a single one; and conversely, the structure being used to describe a single world in the ontological Deutsch-Hayden picture is one we know in fact to be adequate to describe a whole set of possible worlds in quantum mechanics. Thus the Deutsch-Hayden picture, taken ontologically, would seem to be extremely, perhaps implausibly, extravagant in the structure it uses to depict a single world. This difficulty, whilst certainly not a knock-down objection to the ontological intrepretation, nonetheless seves to highlight some of its unpalatable features.

\section{Information and Information Flow}\label{info}

We have seen that under the conservative interpretation, the Deutsch-Hayden formalism does not confer any benefits with respect to locality that do not follow directly from adopting no-collapse, unitary, quantum mechanics as a basic theory, and hence would be equally available with an Everettian interpretation, or, if one were perhaps to allow a formal collapse, but deny that it corresponded to any real process, on a statistical interpretation. With the ontological interpretation, by contrast, we do find something new, but this is better characterized as concerning the reducibility of global properties to local intrinsic properties of subsystems, rather than being a question of locality or nonlocality.

One of the most important aspects of the Deutsch-Hayden approach, however, is the claim that their formalism finally clarifies the nature of information flow in quantum systems; indeed, that it reveals that information can be seen to be transported locally in quantum systems, the phenomena of entanglement assisted communication notwithstanding. It is to this question that we now turn. Again, the matter must be assessed independently for the two different modes of interpretation of the formalism. We shall begin, however, with a few general remarks about the topic of information flow.   

\subsection{Whereabouts of information}\label{whereabouts}

The puzzle that seems to be posed by the examples of teleportation and the like is over the question `How does the information get from $A$ to $B$?'. This is a perfectly legitimate question if it is understood as a question about what the causal processes involved in the transmission of the information are\footnote{Note that different answers to this question will be given depending on one's interpretation of quantum mechanics. For example, in an interpretation involving genuine collapse, action-at-a-distance will play a crucial r\^ole.}, but note that it would be a mistake to take it as a question concerning how information, construed as a particular, or as some pseudo-substance, travels. `Information' is an abstract noun and doesn't serve to refer to an entity or substance. Thus when considering an information transmission process, one that involves entanglement or otherwise, we should not feel it incumbent upon ourselves to provide a story about how some thing, denoted by `the information', travels from $A$ to $B$; nor, \textit{a fortiori}, worry about whether this supposed thing took a spatio-temporally continuous path or not. By contrast, we might very well be interested in the behaviour of the \textit{physical systems} involved in the transmission process and which may or may not usefully be said to be information carriers during the process.    

A second general point concerns what it might mean to ask whether or not information is a `non-local quantity' \cite[p.1759]{dh}. 
Note that for the reason just stated, information is not something that can be said to have a spatio-temporal character, but nonetheless one can, in certain contexts, intelligibly ask `Where is the information?' This question is a fairly specialised one, though: it presupposes that we have some specific piece, or type, of information in mind and asks where this may be found, in the sense of asking where one might learn, or learn about, the fact, or facts, it pertains to. (And, of course, to specify where something may be learnt is not to say that \textit{what is learnt} has to be located there.) Sometimes no very precise answer to this question in terms of a designated spatio-temporal region will be possible, or particularly helpful. 

As a particular example of the latter case, and one that will figure again later, consider the following scenario of encrypting a message. Let us say that Alice and Bob are spatially separated but share a secret random bit string, the key. Alice also has in her possession a message she wishes to send to Bob, a string of bits denoting something; this is the information we are interested in. At this stage, we can say that Alice's notebook, in which the message is written, contains the information. If she then encrypts the message by adding (mod 2) the message string to the key, writes the result down (producing the cyphertext); and destroys both the original message and her copy of the key, then the question `Where is the information now?' leaves us without a straightforward answer. We can't answer by gesturing to Alice's side, or to Bob's side, or to the cyphertext, since from none of these, taken individually, may we learn what the message was; although if we had access both to Bob's key and the cyphertext then we should be able to learn it. A simple request for a location doesn't have a useful answer in this scenario. For this reason, we introduce further vocabulary and talk instead of the message being \textit{encrypted} in the cyphertext. It is not to be found wherever the cyphertext is located, rather, it may be learnt whenever cyphertext and key are brought together, and not otherwise; the asymmetry in the r\^oles of the cyphertext and key is captured by the fact that it is the cyphertext and not the key in which the message is said to be encrypted (although not located).
The bald question `where is the information throughout this protocol?' does not, in this case, invite answers with sufficient articulation for a perspicuous description of what is going on.

Deutsch and Hayden, however, have something specific in mind when they raise the question of whether in quantum systems, information is a local or non-local quantity. If it is the case that a joint quantum system can have global properties that are not reducible to local properties of subsystems, then these global properties might be used to encode and transmit information in a way that cannot be understood as subsystems individually carrying the information. This is what they would mean by information being a non-local quantity. 
The issue is whether we can, in general, always understand an information transmission process involving quantum systems in terms of the properties of subsystems being used to carry the information.
The examples of entanglement assisted communication, as usually understood, would strongly suggest otherwise.

We shall focus on teleportation as the most interesting case; and one which displays the characteristic features at issue.

\subsection{Explaining information flow in teleportation: Locally accessible and inaccessible information}\label{explaining}
 
\begin{figure}
\scalebox{0.7}{\includegraphics{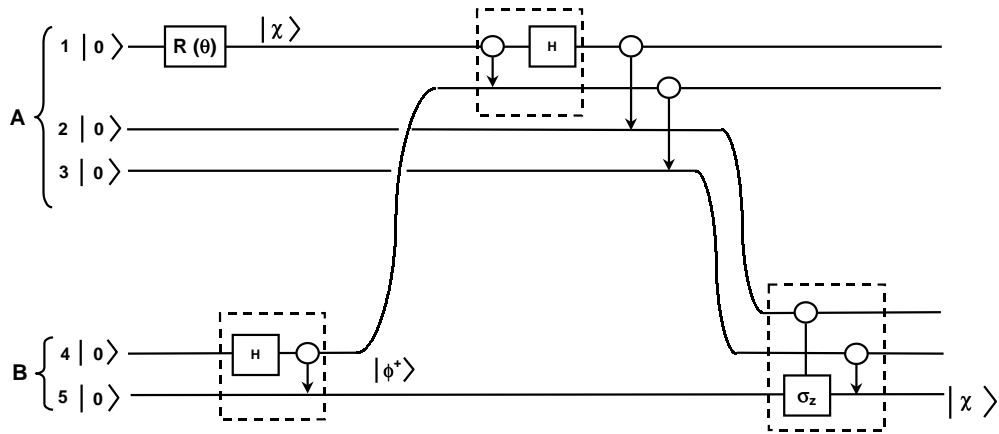}}
\caption{Teleportation. All systems begin in the 0 computational basis state. Bob (B) creates a maximally entangled state of systems 4 and 5. System 1 is prepared in some unknown state $\ket{\chi}{}$, by a rotation depending on the parameter $\theta$. When system 4 is sent to Alice (A), she performs a measurement in the Bell basis, recording the outcome in systems 2 and 3. Systems 2 and 3 are transported to Bob, who performs a controlled-$\sigma_{z}$ operation on 2 and 5,and a controlled-NOT on 3 and 5. System 5 is left in the original unknown state $\ket{\chi}{}$.}\label{teleportation}
\end{figure}

Let us recall once more what the teleportation protocol looks like in the absence of collapse (Fig.~\ref{teleportation}).
Sharing a maximally entangled state with Bob, Alice performs a joint measurement on her half of the entangled pair (4) and on a system (1) prepared in some unknown state, with the result that the state of Bob's system (5), relative to the outcomes of her measurement, is changed in a way that relates systematically to the unknown state to be teleported. At this stage of the protocol, every system involved is now in a maximally mixed state, i.e., the information that characterises the unknown state will not be available to local measurements\footnote{This would not in general be the case if the initial entangled state were not maximally entangled, or if Alice's measurement were not a perfect von Neumann measurement; with these eventualities, the teleportation would be imperfect (fidelity less than 1).}.
As we have seen, the protocol continues with the sending of the systems (2 and 3) recording the outcome of Alice's measurement to Bob, who can now perform the conditional unitary operations required to disentangle his system (5) from the others, in such a way that it ends up in the original, unknown, state. The information characterising the unknown state is now available again to local measurements, but this time, only at Bob's location.  

The crucial feature in this protocol is the change in the relative states that is allowed by the \textit{global} property of entanglement. Subsystems, therefore, do not seem to be playing the r\^ole of information carriers in teleportation, and this conclusion is further supported by the fact that the only systems that are sent from Alice to Bob during the protocol are both maximally mixed.

Deutsch and Hayden, though, wish to give an account of teleportation in which information flow is local; that is, in which subsystems can indeed be seen to carry information from Alice to Bob. In particular, they are concerned to rebut claims such as that of Braunstein \cite{braunstein:irreversible}, who suggests that the information characterising the unknown state is contained in the global system rather than in subsystems during the protocol; or that of Penrose \cite{penrose}, who suggests that the information must flow along a channel constituted by the initial shared entanglement between Alice and Bob, first backwards, and then forwards again in time\footnote{Note that the feeling that an explanation of this somewhat extravagant form is required would seem to be motivated by a picture in which information is a particular, or substance; which, it has been suggested, is misplaced.}. 

Clearly, a good starting point for the debate would be an appropriate criterion for when a system may be said to contain information. 
Deutsch and Hayden would seem to have one of two slightly different necessary and sufficient conditions in mind, although they are not explicit.

They begin by introducing a fairly familiar \textit{sufficient} condition for a system $S$ to contain information about a parameter $\theta$: If a suitable measurement on $S$ would display a probabilistic dependence on $\theta$, then $S$ may be said to contain information about $\theta$. Then a \textit{necessary} condition for containing information is presented: $S$ can be said to contain information about $\theta$ only if its descriptor depends on $\theta$. These definitions motivate an informal argument of roughly the following form: Let us say we have a group of systems that includes $S$; denote this group by $S \cup S^{\perp}$. Assume that the descriptor of $S$ alone depends on $\theta$. If we know that the group $S \cup S^{\perp}$ as a whole contains information about $\theta$, because global measurements would display suitable probabilistic dependence, but $S^{\perp}$ does not (as the descriptors of the systems in $S^{\perp}$ do not depend on $\theta$), then the information must be in $S$, in virtue of $S$'s descriptor depending on $\theta$. Therefore from the fact that the descriptor of $S$ depends on $\theta$, we may infer that it contains information about $\theta$.

This conclusion would be underwritten by either one of the following two definitions\footnote{The two statements that follow must be understood as proposed definitions, as they are not entailed by Deutsch and Hayden's argument, just sketched. The argument uses the necessary and the sufficient condition for containing information, and the rule of inference: if a group of systems contains information about $\theta$, and a subgroup does not, then the complement of that subgroup contains the information about $\theta$. However if we have more than one system whose descriptor depends on $\theta$, then all that the argument based on these principles allows us to conclude is that their union contains the information, not each system individually, which is the desired conclusion.}:
\newtheorem{definition}{Definition}
\begin{definition}
$S$ contains information about $\theta$ $\leftrightarrow$ its descriptor depends on $\theta$
\end{definition}
\begin{definition}
$S$ contains information about $\theta$ $\leftrightarrow$ its descriptor depends on $\theta$ and measurements on the global system $S \cup S^{\perp}$ would display a probabilistic dependence on $\theta$.   
\end{definition}
These two definitions differ as it is possible for the $\mathbf{q}_{i}(t)$ to depend on $\theta$, but for $\rho(t)$ not to (recall the problem of underdetermination). The second is rather more natural, particularly if we are to tie the notion of information being used to the context of definite communication-theoretic procedures.

With one of these definitions of containing information in hand, Deutsch and Hayden's claim for the locality of information flow follows directly from the contiguity property of the changes in the $\mathbf{q}_{i}(t)$. The proposal is that teleportation should now be understood in the following way. System 1 is prepared in some state characterised by the parameter $\theta$; its descriptor now depends on $\theta$. Following Alice's Bell-basis measurement, the descriptors of the `message qubits' 2 and 3 also come to depend on $\theta$. These two systems, as they are transported, carry the information about $\theta$ to Bob's location, where, following a suitable local interaction, the descriptor of his system (5) also comes to depend on $\theta$. We must note the further, crucial, point, however, that the systems 2 and 3 carry the information to Bob in a \textit{locally inaccessible} manner. Although their descriptors depend on $\theta$, and hence the systems may be said to carry information under the Deutsch-Hayden definition, this dependence may not be revealed by measurements on the systems individually --- their reduced density matrices are maximally mixed.

Deutsch and Hayden define locally inaccessible information as information that is present in a system, but that may not be revealed by individual measurements on the system. The explanation of teleportation, then, is that the message qubits do actually carry the information characterising the unknown state to Bob, but they do so locally inaccesibly. The general conclusion is that subsystems can always be thought to carry information in entanglement assisted communication protocols (hence `information is a local quantity'), it is just that these protocols involve locally inaccessible information.  

\subsection{Assessing the claims for information flow}\label{assessinfo}

How satisfactory is this account as an explanation of teleportation, and, indeed as a general picture for information transmission in quantum systems? We shall consider three questions: First, have Deutsch and Hayden finally given the correct account of teleportation, as opposed, say, to Braunstein? Second, is the concept of locally inaccessible information useful? Third, do Deutsch and Hayden provide us with a new concept of information, or quantum information? 
We must consider the answers to these questions for the two modes of interpretation of the formalism in turn.

Before that, a preliminary remark. Recall that as properly understood, the question `How does information get from Alice to Bob?' is a question about the causal processes involved in the transmission. It is clear that simply answering `the information is carried in the message qubits' would not be enough to explain teleportation on its own, as this information might never be made accessible again at Bob's location, or it might be made locally accessible, perhaps, but not in such a way that Bob's system is actually to be found in the original unknown state. Obviously, the explanation has also to refer to the r\^ole of the initial entanglement and the changes in the global properties of the system that this entanglement allows, and which the teleportation protocol exploits. This suggests a moderate way of understanding the application of the Deutsch-Hayden formalism in teleportation that would not involve commitment to their claims about locality or information flow.

On this view, the advantage their formalism presents is simply in highlighting the difference in r\^oles played by the initial entanglement and the message qubits in teleportation. The asymmetry in these r\^oles is, as Deutsch and Hayden point out, analogous to the asymmetry in the r\^oles of the key and cyphertext in classical encryption based on a shared secret random string\footnote{The analogies and, importantly, disanalogies, between entanglement and shared secret bits are developed in detail in \cite{collins:popescu}.}. Before the final stage of the protocol, it is the message qubits, and not Bob's qubit, that have had the direct dynamical coupling to the system whose state is to be teleported (reflected in the fact that their descriptors depend on $\theta$) --- compare with the classical cyphertext, which is generated from the message. But it is the correlations that are established between the relative states of the message systems and Bob's qubit, in virtue of the initial entanglement, that allow the unknown state to be recovered by Bob. (Similarly, the classical correlations between the key and cyphertext allow the encrypted message to be recovered). This suggests that it may well be useful to distinguish between the question of whether an analysis in terms of the $\mathbf{q}_{i}(t)$ helps us understand an aspect of teleportation; and whether the account in terms of information flow does so.      

Returning to our three questions. The adjective `correct' in the first question might be understood in one of two ways; either correct \textit{simpliciter}, or correct \textit{given} the background assumptions. In order to be correct \textit{simpliciter}, the account of teleportation would clearly have to be, first of all, correct given the background assumptions, while these background assumptions themselves also have to be correct. The relevant background assumption when we consider the conservative interpretation is that unitary (no-collapse) quantum mechanics is our setting; this is the setting also for Braunstein \cite{braunstein:irreversible}, hence the point of the comparison.  

\subsubsection{Conservative interpretation}

From the previous remarks on the conservative interpretation, we know that the assignment of properties to systems involves both the global state and the $\mathbf{q}_{i}(t)$: we do not have reducibility of global properties to properties of subsystems and therefore subsystems cannot, after all, always be thought to carry information in entanglement assisted communication. It makes no odds whether one adopts the Heisenberg or the Schr\"odinger viewpoint, it is still the case that joint (and irreducible) properties of subsystems are being used to carry information in the protocols. In Braunstein's account of teleportation, after Alice's Bell-basis measurement, the information characterising the unknown state is said to be in the correlations between the message qubits and Bob's qubit, i.e., it is carried by certain joint properties of these systems. The same is true in the Deutsch-Hayden setting, understood conservatively; so we are not in fact being offered a substantially different account of teleportation. This entails part of the answer to the second question.

Under the conservative interpretation, there is an important sense in which there is no difference between saying that a system contains locally inaccessible information and saying that the information is in the correlations. In both cases this would translate into: the information is carried by joint, and not individual, properties of subsystems. One can frequently make perfectly good sense of a system being said to contain information about a parameter if a suitable measurement on the system would display a probabilistic dependence on the parameter, for then one can learn something about the parameter by performing the measurement. But if the information is locally inaccesible, then this means either i) for some \textit{different} initial state of the global system then there will be a probabilistic dependence for the local measurement --- but this would be physically irrelevant to the situation actually being considered; or ii) for some measurement on the \textit{global} system, a probabilistic dependence on the parameter will be displayed --- and this is no different from what one would say on Braunstein's account.

So where, if anywhere, does a difference lie? In marking an asymmetry. But note that the pertinent aysmmetry may also be understood in a Schr\"odinger picture account such as Braunstein's. In teleportation, the point being emphasized is that it is the message qubits, and not Bob's qubit, that have had the direct dynamical coupling to the system that was prepared in the state characterized by the parameter $\theta$; and this is clear enough without invoking locally inaccessible information. (The significance, of course, is that we know from the no-signalling theorem that dependence on a parameter chosen in one region may not be displayed in another unless there has been a direct, or indirect, dynamical coupling between systems from the two regions.) Another way to mark the asymmetry would begin by pointing out that the initial entanglement, the sending of the message qubits to Bob, and the correct sequence of unitary operations being performed by Alice and Bob, are individually necessary, and jointly sufficient conditions for a successful teleportation protocol. If we were to miss any one of these out, then the protocol would fail, but evidently, for different reasons in each case.         

The preceding discussion indicates that under the conservative interpretation, the concept of locally inaccessible information is not playing a very useful explanatory r\^ole. It is misleading to suggest that the message qubits really carry anything --- at best this is a roundabout way of saying that joint properties do\footnote{Recall from the comments in Section~\ref{whereabouts} that we are not \textit{forced} to say that the information must be located in one system rather than another, or that it is carried by one system rather than another. The assumption that we \textit{must} is predicated upon the misleading picture of information as a particular or substance.}. 
This conclusion in turn casts doubt on the value of adopting either of the proposed definitions of containing information in the context of the conservative interpretation.

However, it would be precipitate to conclude from this that we may in fact learn nothing from the analysis of teleportation in the Deutsch-Hayden formalism. As suggested earlier, one can distinguish between the description using the $\mathbf{q}_{i}(t)$ being useful and the concept of locally inaccessible information being so. Deutsch and Hayden are certainly right that an analysis in terms of their descriptors does help emphasize the important asymmetry between the r\^oles in the protocol of sending the message qubits and the existence of the initial entanglement; and due consideration of this asymmetry contributes, for example, towards undermining the plausibility of a Penrose-type explanation. The analogy with the cyphertext and key is also enlightening in this regard. But as we have just noted, it is quite possible to mark this asymmetry without needing to invoke talk of containing information, which has potential to mislead.

The answer to the third question under the conservative interpretation is perhaps the most intriguing. We have seen that locally inaccessible information does not figure successfully in an attempt to retain subsystems as information carriers in the presence of entanglement, but have Deutsch and Hayden nonetheless succeeded in shedding light on the---sometimes obscure seeming---concept of quantum information? They say, for example:
\begin{quote} 
...it is impossible to characterize quantum information at a given instant using the state vector alone. To investigate where information is located, one must also take into account how the state came about. In the Heisenberg picture this is taken care of automatically, precisely because the Heisenberg picture gives a description that is both complete and local. \cite[p.1773]{dh}
\end{quote}
It seems, though, that this suggestion would incorporate a number of confusions.

While it is true that the $\mathbf{q}_{i}(t)$ provide more information than simply following the time evolved state would, this is not information about the time evolution of particular systems that the latter description lacks. The $\mathbf{q}_{i}(t)$ look more informative because they capture time evolution for any given initial state, thus they say more about the dynamics a system has been subject to; but in the conservative interpretation, this is not to say more about the system, but rather about the \textit{unitary operators}.
This extra information that one gets is not then `complete', i.e., information that would be lacking in the description of a given network of systems in the Schr\"odinger picture, but is given one in Heisenberg. Instead, it is information about something else; about how other systems, prepared in a different way would react, or information about, for example, the fields that have driven the systems' evolution.

Furthermore, one can readily accept that one has more information if one knows how the state came about, but deny that this information is a property that has to be located. So again, one can, in fact should, deny that there is information located with systems that is lacking from the state vector picture. The `extra' information represented in the $\mathbf{q}_{i}(t)$ consists of facts about the unitary operations undergone; and this information cannot be said to be here, there, or anywhere, as it makes no sense to ask where these facts are. Facts are of the wrong logical category to possess a location (cf.~\cite{strawson}). The underlying thought seems to be that the description in terms of the $\mathbf{q}_{i}(t)$ allows us to `determine where the information about a given parameter is located at a given instant' \cite[p.1771]{dh}. But note that the question `Where is the dependence on the parameter?' could be a bad question; one inviting us to confuse the description of a thing with the thing itself. It is \textit{what} depends on the parameter that is important; and in entanglement assisted communication, under the conservative interpretation, this will often only be \textit{joint}, and not individual properties.   


\subsubsection{Ontological interpretation}

The discussion of our three questions for the ontological interpretation may be somewhat more brief. As to the first: on the ontological interpretation, global properties are reduced to intrinsic properties of subsystems, therefore, the properties of subsystems may indeed be thought to be carrying the information in entanglement assisted communication protocols. Thus, adopting the Deutsch-Hayden formalism understood in the ontological way, we would have an explanation of teleportation in which
the information that the system carries as a whole can be thought a consequence of information being carried by subsystems; in which information is genuinely carried between Alice and Bob in the message qubits during teleportation. (Of course, this explanation may not be reflected back onto our more usual ways of understanding quantum mechanics, but relies on the ontological interpretation. As such it has no power to confute opposing views, such as Braunstein's, that derive from a different set of assumptions.) 


Why does it now seem acceptable to say that information is carried in subsystems, despite the fact that it may not be possible to learn anything by performing measurements on an individual system? Because in the ontological interpretation, the explanation of the physical processes by which information is transmitted from $A$ to $B$ (answering `How does the information get from $A$ to $B$' in the legitimate way,) involves the intrinsic properties of subsystems denoted by the $\mathbf{q}_{i}(t)$.
In contrast to the conservative interpretation, we are now able to answer the question `What depends on the parameter?' with: the intrinsic properties of subsystems. 
As the intrinsic properties of subsystems are being used as the information bearing properties under the ontological interpretation, the definitions given above of containing information would have a point\footnote{Although it is not clear that they are wholly trouble-free. Under definition (1), for example, there will be cases in which a system is said to contain information locally inaccessibly, but where it could never be made accessible, i.e. could never be displayed even under global measurements. This would tend to undermine the plausibility of the claim that the system does in fact contain information, which casts doubt on the acceptability of the definition. So again, definition (2) would seem preferable. But  
it might be beneficial to restrict talk of containing information still further, to cases in which some particular information transmission protocol is envisaged, or in which an agent would stand to learn something by performing measurements on a group of systems. 
}.
 
Regarding the usefulness of the concept of locally inaccessible information, the purpose of the introduction of this category is to recognise that there are two ways in which a system may be said to carry information in the ontological interpretation; either in its observable, or in its unobservable, empirically inaccessible, properties. This distinction is necessary for the explanation of entanglement assisted communication in the ontological interpretation, thus the introduction of the category is useful.

In answer to our third question, however, it is important to recognise that the ontological interpretation of Deutsch-Hayden is not providing us with an account of a new type of information, but of new properties, new ways in which information may be carried. Again, because this turns on the details of the ontological interpretation, it cannot be taken to provide us with a new understanding of information, or quantum information, that could be transferred back to more familiar quantum mechanical settings.

\section{Conclusion}

Deutsch and Hayden present their formalism as an avowedly local account of quantum mechanics, which finally clarifies the nature of information transmission in entangled quantum systems. To what extent is this successful? We have seen that in order to assess the claims of locality, and the claims regarding the nature of information flow, it is essential to distinguish between a conservative and an ontological interpretation of the formalism, as very different conclusions follow. To summarise:

On the conservative interpretation, there are no benefits with respect to locality that do not follow immediately from adopting a version of quantum mechanics in which there is no genuine process of collapse and no additional properties added (and which, consequently, would be shared by an Everettian or a statistical interpretation); thus no distinctive feature of the Deutsch-Hayden approach is in play. As far as information transmission is concerned, the formalism does not show that information is after all, a local quantity (in Deutsch and Hayden's sense), as it remains the case that joint, rather than individual, properties are used to carry information in entanglement assisted communication protocols. The explanation proffered of teleportation does not differ in substance from that which would be given by an account sharing the same initial assumptions, such as that of Braunstein. Furthermore, we have seen that it would be confused to think that the description in terms of the $\mathbf{q}_{i}(t)$ fills-in an account of information, and where it is located in quantum systems, that is missing in the usual Schr\"odinger picture. The additional information the $\mathbf{q}_{i}(t)$ provide (when they do so) consists of certain facts about the unitary operations undergone (not information carried by systems); and it makes no sense to propose that these facts have a location. 

With the ontological interpretation, on the other hand, we have an interesting result; although one better characterized as regarding the reducibility of global properties of quantum systems to individual properties, rather than as a question of locality or nonlocality. With this reducibility, the claim about the locality of information transmission, even in the presence of entanglement, follows. However, as the ontological interpretation provides a picture which differs so markedly from our usual ways of understanding quantum mechanics, these results clearly cannot be taken to shed light on the nature of information flow in entangled quantum systems when we have \textit{not} taken the dramatic step of introducing an entirely new range of intrinsic properties of systems. And reducibility does not come free: one is confronted with an unpleasant form of underdetermination and the bogey of redundancy.   

Unfortunately, Deutsch and Hayden do not distinguish the two different modes of interpretation of their formalism; indeed they are arguably conflated, to deleterious effect. The reason to believe that they must have something along the lines of the ontological interpretation in mind is that their main claims would not be true in any interesting way otherwise; but at certain points they would seem to suggest clearly that the conservative reading is correct: when they imply that it is merely the move to the Heisenberg picture which does the work (p.1759); when suggesting that they have simply provided a reformulation of Schr\"odinger picture quantum mechanics (p.1773).
As we have seen, however, if there is equivocation between the conservative and the ontological interpretation, then it is impossible to draw any conclusion regarding information flow and locality.

So, having drawn this all-important distinction, the conclusion of our discussion is that in the ontological interpretation, we have a bold thesis which might be adopted, despite its objectionable features, in order to obtain reducibility of global properties to local properties, if this was thought particularly desirable for some reason. Retaining the conservative approach, on the other hand, we would have a formalism with some occasionally useful features, but not one which provides a novel sense of locality, nor, indeed, of information flow. \textit{En route}, the discussion has hopefully shed some light on the puzzles that so often seem to surround the question of information transmission in entangled quantum systems.

\section*{Acknowledgements}
Thanks are due to Jon Barrett, Harvey Brown and Richard Jozsa for useful discussion of the topics raised here. My understanding of these matters has benefited considerably from numerous conversations with David Wallace. Thanks also to Jane Timpson for drawing the figures. This work was partially supported by a studentship from the UK Arts and Humanities Research Board.


\end{document}